\documentclass[]{aa} 
\usepackage{graphicx}
\usepackage{txfonts}
\usepackage{natbib}
\usepackage{ulem}
\usepackage{url}
\usepackage[usenames]{color}
\bibpunct{(}{)}{;}{a}{}{,} 
\newcount\longrefs
\def\aap{\ifnum\longrefs=1 {Astron.\ Astrophys.}\else 
                           {A\hbox{\rm \&}A}\fi}
\def\aapr{\ifnum\longrefs=1 {Astron.\ Astrophys.\ Rev.}\else 
                            {A\hbox{\rm \&}AR}\fi}
\def\aaps{\ifnum\longrefs=1 {Astron.\ Astrophys.\ Suppl.}\else 
                            {A\hbox{\rm \&}A Suppl.}\fi}
\def\aj{\ifnum\longrefs=1 {Astron.\ J.}\else 
                          {AJ}\fi} 
\def\ao{\ifnum\longrefs=1 {Applied Optics}\else 
                           {Appl.\ Opt.}\fi} 
\def\aspcs{\ifnum\longrefs=1 {Astron.\ Soc.\ Pacific Conf. Series}\else 
                           {ASP Conf.\ Ser.}\fi} 
\def\apj{\ifnum\longrefs=1 {Astrophys.\ J.}\else 
                           {ApJ}\fi} 
\def\apjl{\ifnum\longrefs=1 {Astrophys.\ J. Lett.}\else 
                            {ApJ}\fi} 
\def\aplett{\ifnum\longrefs=1 {Astrophys.\ J. Lett.}\else 
                            {ApJ}\fi} 
\def\apjs{\ifnum\longrefs=1 {Astrophys.\ J. Suppl.}\else 
                            {ApJS}\fi}
\def\apss{\ifnum\longrefs=1 {Astrophys.\ and Space Science}\else 
                            {Astrophys.\ Space Sci.}\fi}
\def\araa{\ifnum\longrefs=1 {Ann.\ Rev.\ Astron.\ Astrophys.}\else 
                            {ARA\hbox{\rm \&}A}\fi}
\def\azh{\ifnum\longrefs=1 {Astronomicheskii Zhurnal}\else 
                            {Astron.\ Zhur.}\fi}
\def\baas{\ifnum\longrefs=1 {Bull.\ Am.\ Astron.\ Soc.}\else 
                            {BAAS}\fi}
\def\bain{\ifnum\longrefs=1 {Bull.\ Astronom.\ Institutes Netherlands}\else
                            {Bull.\ Astr.\ Inst.\ Neth.}\fi}
\def\gca{\ifnum\longrefs=1 {Geochim.\ Cosmochim.\ Acta}\else 
                           {Geochim.\ Cosmochim.\ Acta}\fi}
\def\grl{\ifnum\longrefs=1 {Geophys.\ Res.\ Lett.}\else 
                           {Geoph.\ Res.\ Lett.}\fi}
\def\iaucirc{\ifnum\longrefs=1 {IAU Circulars}\else 
                          {IAU Circ.}\fi}
\def\ip{\ifnum\longrefs=1 {in press}\else 
                          {in press}\fi}
\def\jgr{\ifnum\longrefs=1 {J.\ Geophys.\ Res.}\else 
                           {J.\ Geophys.\ Res.}\fi}  
\def\jrasc{\ifnum\longrefs=1 {J.\ Royal Astron.\ Soc.\ Canada}\else 
                           {JRAS Can.}\fi}  
\def\mnras{\ifnum\longrefs=1 {Mon.\ Not.\ Roy.\ Astron.\ Soc.}\else 
                             {MNRAS}\fi} 
\def\nat{\ifnum\longrefs=1 {Nature}\else 
                           {Nat}\fi}
\def\pasj{\ifnum\longrefs=1 {Pub.\ Astron.\ Soc.\ Japan}\else 
                            {PASJ}\fi} 
\def\pasp{\ifnum\longrefs=1 {Pub.\ Astron.\ Soc.\ Pacific}\else 
                            {PASP}\fi} 
\def\physscr{\ifnum\longrefs=1 {Physica Scripta}\else 
                            {Phys.\ Scrip.}\fi} 
\def\planss{\ifnum\longrefs=1 {Planetary \& Space Science}\else 
                            {Plan. \& Space Sci.}\fi} 
\def\procspie{\ifnum\longrefs=1 {Proc.\ SPIE}\else 
                            {Proc.\ SPIE}\fi} 
\def\qjras{\ifnum\longrefs=1 {Quarterly J.\ Royal Astron.\ Soc.}\else 
                            {QJRAS}\fi} 
\def\sa{\ifnum\longrefs=1 {Soviet Astron..}\else 
                               {Sov.\ Astron.}\fi}
\def\skytel{\ifnum\longrefs=1 {Sky \& Telescope}\else 
                            {Sky \& Tel.}\fi} 
\def\solphys{\ifnum\longrefs=1 {Solar Phys.}\else 
                               {Sol.\ Phys.}\fi}
\def\ssr{\ifnum\longrefs=1 {Space Science Rev.}\else 
                               {Space\ Sci.\ Rev.}\fi}

\def\nl{,\ } 

\def\CMAO{Center of Mathematics for Applications, University of Oslo\nl
           P.O. Box 1053, Blindern\nl N-0316 Oslo\nl Norway}

\def\Oslo{Institute of Theoretical Astrophysics, University of Oslo\nl 
         P.O. Box 1029, Blindern\nl N--0315 Oslo\nl Norway}

\def\SIU{Sterrekundig Instituut, Utrecht University\nl Postbus 80\,000\nl
         NL--3508 TA Utrecht\nl The Netherlands}

\def\dutch{\def\refname{Referenties}\def\abstractname{Samenvatting}%
  \def\bibname{Bibliografie}\def\chaptername{Hoofdstuk}%
  \def\appendixname{Bijlage}\def\contentsname{Inhoudsopgave}%
  \def\listfigurename{Lijst van figuren}%
  \def\listtablename{Lijst van tabellen}%
  \def\indexname{Index}\def\figurename{Figuur}\def\tablename{Tabel}%
  \def\partname{Deel}\def\enclname{Bijlage(n)}\def\ccname{Ter attentie van}%
  \def\headtoname{Aan}\def\headpagename{Pagina}%
  \def\today{\number\day\space\ifcase\month\or januari\or februari\or%
     maart\or%
     april\or mei\or juni\or juli\or augustus\or september\or oktober\or%
     november\or december\fi \space\number\year}%
  \typeout{
              >>>>> use hlatex209 for Dutch hyphenation <<<<< 
         }}

\DeclareFontFamily{OT1}{mvs}{}
\DeclareFontShape{OT1}{mvs}{m}{n}{<-> fmvr8x}{}

\newcounter{onefig} \newcounter{fignumber}
\newcount\nocaptions \newcount\nofigures \newcount\figwidth
\newcount\viewgraphs \newcount\printlabel
\long\def\skipfigure #1\viewout{}   
  \def\paper{}  \def\figlabel{} 
\long\def\nextfig#1{\setcounter{figure}{\value{fignumber}}
  \addtocounter{fignumber}{1}
  \ifnum \viewgraphs=1 \pagestyle{empty} \fi 
  \ifnum\value{onefig}=0 #1 \fi                 
  \ifnum\value{onefig}=\value{fignumber} #1 \fi}
\def\figwidths#1#2{\ifnum \nocaptions=1 #2mm \else #1mm \fi}  
\def\picplace{\framebox[80mm]{\rule{0cm}{1cm}}}
\def\paper#1{}  
\long\def\plotfig#1#2{\ifnum \nofigures=1 \picplace \else #2 \fi}
\long\def\captiontext#1{\ifnum \nofigures=1 \raggedright \fi 
   \ifnum \nocaptions=1 \paper
     \ifnum \viewgraphs=0 
       \newline  \mbox{}\hrulefill\mbox{} \newline 
       \ifnum \printlabel=1 \{{\em \figlabel}\}\newline \fi
     \fi 
   \else \ifnum \printlabel=1 \{{\em \figlabel}\}\newline \fi
     #1 \fi}

\newcount\panelwidth \newcount\panelheight 
\newcount\bxmin \newcount\bymin \newcount\bxmax \newcount\bymax
\newcount\tbxmin \newcount\tbymin
\newcount\tpanelwidth \newcount\tpanelheight \newcount\tpdif
\def\panelsize #1,#2;{\panelwidth=#1 \panelheight=#2}  
\def\setbb #1,#2;#3,#4;#5,#6;{
  \tbxmin=#1 \tbymin=#2    
  \bxmin=#3 \bymin=#4      
  \bxmax=#5 \bymax=#6}     
\def\barepanel #1{%
  \ifnum\panelheight=0 
    \tpdif=\bymax \advance\tpdif by -\bymin
    \multiply \tpdif by \panelwidth
    \tpanelheight=\tpdif
    \tpdif=\bxmax \advance\tpdif by -\bxmin
    \divide \tpanelheight by \tpdif
  \else \tpanelheight=\panelheight \fi
  \ifnum\panelwidth=0 
    \tpdif=\bxmax \advance\tpdif by -\bxmin
    \multiply \tpdif by \panelheight
    \tpanelwidth=\tpdif
    \tpdif=\bymax \advance\tpdif by -\bymin
    \divide \tpanelwidth by \tpdif
  \else \tpanelwidth=\panelwidth \fi
  \epsfig{file=#1,silent=,%
     bbllx=\bxmin bp,bblly=\bymin bp,bburx=\bxmax bp,bbury=\bymax bp,clip=,%
     width=\tpanelwidth mm,height=\tpanelheight mm}}
\def\labelypanel #1{
  \ifnum\panelheight=0 
    \tpdif=\bymax \advance\tpdif by -\bymin
    \multiply \tpdif by \panelwidth
    \tpanelheight=\tpdif
    \tpdif=\bxmax \advance\tpdif by -\bxmin
    \divide \tpanelheight by \tpdif
  \else \tpanelheight=\panelheight \fi
  \ifnum\panelwidth=0 
    \tpdif=\bxmax \advance\tpdif by -\bxmin
    \multiply \tpdif by \panelheight
    \tpanelwidth=\tpdif
    \tpdif=\bymax \advance\tpdif by -\bymin
    \divide \tpanelwidth by \tpdif
  \else \tpanelwidth=\panelwidth \fi
  \tpdif=\bxmax \advance\tpdif by -\tbxmin
  \multiply \tpanelwidth by \tpdif
  \tpdif=\bxmax \advance\tpdif by -\bxmin
  \divide \tpanelwidth by \tpdif
  \epsfig{file=#1,silent=,%
    bbllx=\tbxmin bp,bblly=\bymin bp,bburx=\bxmax bp,bbury=\bymax bp,%
    clip=,width=\tpanelwidth mm,height=\tpanelheight mm}}
\def\labelxpanel #1{%
  \ifnum\panelheight=0 
    \tpdif=\bymax \advance\tpdif by -\bymin
    \multiply \tpdif by \panelwidth
    \tpanelheight=\tpdif
    \tpdif=\bxmax \advance\tpdif by -\bxmin
    \divide \tpanelheight by \tpdif
  \else \tpanelheight=\panelheight \fi
  \ifnum\panelwidth=0 
    \tpdif=\bxmax \advance\tpdif by -\bxmin
    \multiply \tpdif by \panelheight
    \tpanelwidth=\tpdif
    \tpdif=\bymax \advance\tpdif by -\bymin
    \divide \tpanelwidth by \tpdif
  \else \tpanelwidth=\panelwidth \fi
  \tpdif=\bymax \advance\tpdif by -\tbymin
  \multiply \tpanelheight by \tpdif
  \tpdif=\bymax \advance\tpdif by -\bymin
  \divide \tpanelheight by \tpdif
  \epsfig{file=#1,silent=,%
    bbllx=\bxmin bp,bblly=\tbymin bp,bburx=\bxmax bp,bbury=\bymax bp,%
    clip=,width=\tpanelwidth mm,height=\tpanelheight mm}}
\def\labelxypanel #1{%
  \ifnum\panelheight=0 
    \tpdif=\bymax \advance\tpdif by -\bymin
    \multiply \tpdif by \panelwidth
    \tpanelheight=\tpdif
    \tpdif=\bxmax \advance\tpdif by -\bxmin
    \divide \tpanelheight by \tpdif
  \else \tpanelheight=\panelheight \fi
  \ifnum\panelwidth=0 
    \tpdif=\bxmax \advance\tpdif by -\bxmin
    \multiply \tpdif by \panelheight
    \tpanelwidth=\tpdif
    \tpdif=\bymax \advance\tpdif by -\bymin
    \divide \tpanelwidth by \tpdif
  \else \tpanelwidth=\panelwidth \fi
  \tpdif=\bxmax \advance\tpdif by -\tbxmin
  \multiply \tpanelwidth by \tpdif
  \tpdif=\bxmax \advance\tpdif by -\bxmin
  \divide \tpanelwidth by \tpdif 
  \tpdif=\bymax \advance\tpdif by -\tbymin 
  \multiply \tpanelheight by \tpdif
  \tpdif=\bymax \advance\tpdif by -\bymin
  \divide \tpanelheight by \tpdif
  \epsfig{file=#1,silent=,%
    bbllx=\tbxmin bp,bblly=\tbymin bp,bburx=\bxmax bp,bbury=\bymax bp,%
    clip=,width=\tpanelwidth mm,height=\tpanelheight mm}}

\def\CC{\par \vspace*{-2ex} \footnotesize \baselineskip=8pt \begin{verbatim}}

\long\def\startignore #1\stopignore{}   

\def\setlistparams{         
  \topsep=0.7ex                 
  \itemsep=0.7ex                
  \leftmargini=3ex}             
\setlistparams                  

\newcounter{alistindex}       

\newcounter{romenumnr}

\newlength{\minipagewidth}

\newsavebox{\boxcontent}
\newcommand{\ovalhead}[1]{
  \unitlength=1cm
  \sbox{\boxcontent}{\mbox{~~{#1}~~}}
  \begin{center}
    \ifdim\wd\boxcontent>6ex 
    \ifdim\wd\boxcontent<8cm 
    \begin{picture}(8,3) \thicklines     
      \put(4.0,0.8){\oval(8,1.6)} 
      \put(0.0,0.7){\parbox{8cm}{
         \begin{center} \usebox{\boxcontent} \end{center}}}
    \end{picture}
    \else \ifdim\wd\boxcontent<12cm 
    \begin{picture}(12,3) \thicklines     
        \put(6.0,0.8){\oval(12,1.6)} 
        \put(0.0,0.7){\parbox{12cm}{
           \begin{center} \usebox{\boxcontent} \end{center}}}
    \end{picture}
    \else
    \begin{picture}(16,3) \thicklines     
        \put(8.0,0.8){\oval(16,1.6)} 
        \put(0.0,0.7){\parbox{16cm}{
           \begin{center} \usebox{\boxcontent} \end{center}}}
    \end{picture}
    \fi \fi \fi
  \end{center}} 

\newcounter{headnr}            
\newcounter{subheadnr}[headnr]
\newcounter{subsubheadnr}[subheadnr]

\font\dropfont= cmr12 scaled \magstep5
\def\dropcap#1#2{{\noindent
    \setbox0\hbox{\dropfont #1}\setbox1\hbox{#2}\setbox2\hbox{(}%
    \count0=\ht0\advance\count0 by\dp0\count1\baselineskip
    \advance\count0 by-\ht1\advance\count0by\ht2
    \dimen1=.5ex\advance\count0by\dimen1\divide\count0 by\count1
    \advance\count0 by1\dimen0\wd0
    \advance\dimen0 by.25em\dimen1=\ht0\advance\dimen1 by-\ht1
    \global\hangindent\dimen0\global\hangafter-\count0
    \hskip-\dimen0\setbox0\hbox to\dimen0{\raise-\dimen1\box0\hss}%
    \dp0=0in\ht0=0in\box0}#2}

\def\rmit#1{{\it #1}}              
           
\def\ie{\rmit{i.e.,}}              
\def\eg{\rmit{e.g.,}}              

\def\specchar#1{\uppercase{#1}}    

\def\CaII{\mbox{Ca\,\specchar{ii}}}

\def\HI{\mbox{H\,\specchar{i}}} 
 
\def\Hmin{\hbox{\rmH$^{^{_{\scriptstyle -}}}$}}      

\def\MgII{\mbox{Mg\,\specchar{ii}}}

\def\Halpha{\mbox{H\hspace{0.1ex}$\alpha$}} 

\def\Lyalpha{\mbox{Ly$\hspace{0.2ex}\alpha$}}

\def\CaIIH{\mbox{Ca\,\specchar{ii}\,\,H}}

\def\HK{\mbox{H\,\&\,K}}

\def\KtwoV{\mbox{K$_{2V}$}}

\def\HtwoV{\mbox{H$_{2V}$}}

\def\level #1 #2#3#4{$#1 \: ^{#2} \mbox{#3} ^{#4}$}   

\def\rmc{{\rm c}}  
\def\rmd{{\rm d}}  
\def\rme{{\rm e}}

 \def\rmH{{\rm H}}

\def\rml{{\rm l}}

 \def\rmW{{\rm W}}

\def\={\hbox{$\!=\!$}}                     

\def\mathstacksym#1#2#3#4#5{\def#1{\mathrel{\hbox to 0pt{\lower 
    #5\hbox{#3}\hss} \raise #4\hbox{#2}}}}

\mathstacksym\lta{$<$}{$\sim$}{1.5pt}{3.5pt} 
\mathstacksym\gta{$>$}{$\sim$}{1.5pt}{3.5pt} 
\mathstacksym\lrarrow{$\leftarrow$}{$\rightarrow$}{2pt}{1pt} 
\mathstacksym\lessgreat{$>$}{$<$}{3pt}{3pt} 

\def\exp{\rme}

\def\HI{\ion{H}{i}}

\def\no{\ensuremath{n_{\mathrm{o}}}}

\def\nel{\ensuremath{n_\rme}}

\def\eo{\ensuremath{e_{\mathrm{o}}}}

\def\eh2{\ensuremath{e_{\mathrm{H2}}}}
\def\nh2{\ensuremath{n_{\mathrm{H2}}}}
\def\exp{\rme}

\def\Hi{\ion{H}{i}}

\def\h2{\ensuremath{\mathrm{H}_2}}

\def\CaII{\ion{Ca}{ii}}
\def\MgII{\ion{Mg}{ii}}

\begin{document}

\title{On the minimum temperature of the quiet solar chromosphere.}

\titlerunning{}
  
\subtitle{}

   \author{J.~Leenaarts \inst{1}
     \and
     M.~Carlsson \inst{2,3}
     \and
     V.~Hansteen \inst{2,3}
     \and
     B.~V.~Gudiksen \inst{2,3}
   }

   \offprints{ J. Leenaarts, \\ \email{j.leenaarts@uu.nl} }

   \institute{ \SIU \and \Oslo \and \CMAO}
   
   \date{Received; accepted}

   \abstract
       {}
   {We aim to provide an estimate of the minimum temperature of the
     quiet solar chromosphere}
{We perform a 2D radiation-MHD simulation spanning the upper
  convection zone to the lower corona. The simulation includes non-LTE
  radiative transfer and {an equation-of-state that includes non-equilibrium ionization of hydrogen and
  non-equilibrium \h2 molecule formation}. We
  analyze the reliability of the various assumptions made in our
  model in order to assess the realism of the simulation.}
{Our simulation contains pockets of cool gas
with down to 1660 K from 1~Mm up to 3.2 Mm height. It
overestimates the radiative heating, and
contains non-physical heating below 1660~K. Therefore we conclude that
cool pockets in the quiet solar chromosphere might have even
lower temperatures than in the simulation, provided that there exist
areas in the chromosphere without significant magnetic heating. 
We suggest off-limb molecular spectroscopy to look for such
cool pockets and 3D simulations including a local dynamo and a
magnetic carpet to investigate Joule heating in the quiet
chromosphere.}  {}

  
   \maketitle

\section{Introduction}                          \label{sec:introduction}

The internetwork solar chromosphere is continuously pervaded by
acoustic waves and shocks with periods of around 3~minutes. A large
literature exists on this subject. Some examples are:
\citet{2001A&A...379.1052K}, 
who studied upper-photospheric and low-chromospheric oscillations
observed in the UV continuum around 150\,nm with the Transition Region
and Coronal Explorer (TRACE).
\citet{2008SoPh..tmp...28R}
studied the response of chromospheric diagnostics to photospheric
events such as granular buffeting and exploding granules as observed with
the Dutch Open Telescope (DOT). Somewhat longer
ago, 
\citet{1993ApJ...414..345L}
studied solar chromospheric oscillations using spectrograms in \CaIIH\ 
 obtained with the Vacuum Tower Telescope at NSO/Sacramento Peak.
\citet{2000ApJ...531.1150W} 
studied upper-chromospheric and transition region oscillations using
spectral time series obtained around 103~nm with the SUMER
spectrograph aboard SOHO. These and other studies confirm the picture
of acoustic wave excitation by granular dynamics in the photosphere,
the waves then propagate predominantly in the vertical direction and
evolve into shocks. Depending on the magnetic field topology they
occasionally retain their identity all the way up into the
transition zone 
\citep{2001ApJ...548L.237M}.
The shock fronts are high temperature disturbances
in an otherwise cool background atmosphere.

Observations of CO lines in quiet sun areas
\citep[\eg][]{1972ApJ...176L..89N,1981ApJ...245.1124A,2006ApJS..165..618A}
indicate temperatures low enough to form significant amounts of CO
molecules. Static 1D semi-empirical models designed to reproduce such
observations can have temperatures down to 2750~K
\citep{1996ApJ...460.1042A}.
These low temperatures in such static models correspond to the cool
inter-shock phases of the quiet chromosphere.

This scenario was confirmed in the one-dimensional (1D) simulations of
\citet{1997ApJ...481..500C} 
that explained the formation of \CaII\ \KtwoV\ and \HtwoV\  bright grains as an effect
of upward-propagating shock waves excited in the photosphere.  The first
attempt to model such wave-response in the chromosphere with
multidimensional numerical simulations was made by 
\citet{2000ApJ...541..468S}.
They studied the response of the chromosphere to collapsing granules
using a 3D radiation-hydrodynamic (RHD) simulation spanning from the upper
convection zone to 1.2~Mm above $ \left< \tau_{500} \right> = 1$, and
again confirmed the picture of the internetwork chromosphere as a cool
background state pervaded by waves and shocks.
\citet{2004A&A...414.1121W}
performed a 3D RHD simulation with higher spatial resolution but
less sophisticated radiative transfer. The higher resolution allowed
them to distinguish different shock front shapes depending on the
excitation mechanism: spherical fronts excited by pressure
fluctuations in intergranular lanes; planar fronts, excited by
simulation-box oscillations, the simulated counterpart of solar
$p$-modes; and the previously described irregular front shapes caused
by collapsing granules. The simulations employed simplified radiative
transfer and an equation of state (EOS) based on Saha ionization equilibrium for hydrogen,
an assumption that fails spectacularly in the chromosphere
\citep{2002ApJ...572..626C}. 
Thus, these 3D simulations properly model the shock compression and
post-shock expansion, but inaccurately incorporate radiative heating,
and give erroneous temperatures from the mass density and internal
energy because of the assumption of Saha ionization equilibrium.


The 1D simulations performed by 
\citet{1997ApJ...481..500C} 
modeled the radiative losses in great detail, including the effect
of slow hydrogen ionization/recombination on the equation of state.
However, the 1D geometry affected the temperatures in the
chromosphere, leading to too high temperatures in the shocks as well
as in the inter-shock phases.
\citet{2007A&A...473..625L} 
performed a 2D radiation-magneto-hydrodynamic (RMHD) simulation with an EOS that took non-equilibrium
hydrogen ionization into account using approximations for the
radiation field based on
\citet{sollum1999}.
This simulation, however, still computed the radiative losses based on
Saha equilibrium, and required an ad-hoc heating term that
prevented the temperature to drop below 2400~K.

All of the above simulations were limited in such a way that an
accurate prediction of the minimum temperature occurring in the quiet
solar chromosphere could not be made.

In this paper we discuss the temperature structure in a simulation
that tries to remedy the various limitations of the previous
models: it is two-dimensional, includes a non-equilibrium
equation of state, includes radiative losses without the assumption of
Saha equilibrium and includes heating through absorption of coronal
radiation. {We do not include a significant magnetic field, so
  that our simulation serves as a baseline against which simulations
  with stronger field can be compared.}

In Sec.~\ref{sec:model} we discuss the RMHD model assumptions in detail
and discuss some of the basic physical processes that occur in the
quiet chromosphere. In Sec.~\ref{sec:results} we discuss the results
of our run, especially paying attention to the accuracy of the
radiative heating. We finish with a discussion and our conclusions in
Sec.~\ref{sec:discussion}.

\section{The model}                          \label{sec:model}

We model the chromosphere using the RMHD code Bifrost
\citep{Gudiksen++2011}.
%

\paragraph{MHD equations.} The Bifrost
code solves the equations of resistive MHD including the effects of
heat conduction and radiation:

\begin{equation} \label{eq:mass}
\frac{\partial \rho}{\partial t} + \nabla \cdot (\rho \vec{u}) = 0,
\end{equation}

\begin{equation} \label{eq:momentum}
\frac{\partial \rho \vec{u} }{\partial t} + \nabla \cdot (\rho \vec{u}
\vec{u} - \uuline{\tau}) = - \nabla P + \vec{j} \times \vec{B} + \rho \vec{g},
\end{equation}

\begin{equation} \label{eq:energy}
\frac{\partial e}{\partial t} + \nabla \cdot (e \vec{u}) = 
- P (\nabla \cdot \vec{u})
- \nabla \cdot \vec{F}_\mathrm{c}  
- \nabla \cdot \vec{F}_\mathrm{r} 
+ Q_\mathrm{visc}
+ Q_\mathrm{Joule},
\end{equation}

\begin{equation} \label{eq:current}
\mu_0 \vec{j} = \nabla \times \vec{B},
\end{equation}

\begin{equation} \label{eq:bfield}
\frac{\partial \vec{B}}{\partial t} = \nabla \times ( \vec{u} \times
\vec{B}) - \nabla \times \uuline{\eta} \vec{j},
\end{equation}
\noindent
with $\rho$ the mass density, $\vec{u}$ the velocity, $\uuline{\tau}$
the viscous stress tensor, $P$ the gas pressure, $\vec{j}$ the current
density, $\vec{B}$ the magnetic field, $\vec{g}$ the gravitational
acceleration, $e$ the internal energy per volume, $\vec{F}_\mathrm{c}$
the energy flux due to heat conduction, $\vec{F}_\mathrm{r}$ the
radiative energy flux, $Q_\mathrm{visc}$ the viscous energy
dissipation, $Q_\mathrm{Joule}$ the Joule heating, $\mu_0$ the
permeability of free space and $\uuline{\eta}$ the electrical
resistivity tensor. Bifrost employs a staggered grid and uses
sixth-order
operators to compute spatial derivatives. The time-stepping is
done using a third-order predictor-corrector scheme by 
\citet{Hyman79}, modified for variable timestep.

\paragraph{Equation of state.} 
Conservation of internal energy is expressed as 
\begin{eqnarray} \label{eq:edist}
e & = & \frac{3 k T}{2} \left(\nel + \nh2 + \sum_{i=1}^{n_{\rml}} n_i + \no
\right)  + \nh2 (\eh2 + \chi_{\mathrm{H}2})  \nonumber  \\
& & + \sum_{i=1}^{n_{\rml}}  n_i \, \chi_i + \eo(\nel,T).
\end{eqnarray}
Here, $k$, $T$, $\nel$, $\nh2$, $n_\rml$, $n_i$, $\no$ are Boltzmann's constant,
the gas temperature, the electron density, the H2 molecule density,
the number of levels in the hydrogen model atom,
the density of atomic hydrogen in excitation or ionization state $i$
and the number density of all other atoms and molecules that are not,
or do not contain, hydrogen, respectively. The rotational and vibrational energy per
\h2 molecule is $\eh2$, $\chi_{\rmH2}$ is the dissociation energy of \h2,
$\chi_i$ is the excitation or ionization energy of atomic hydrogen and
$\eo$ is the internal energy of all other atoms and molecules that are
not, or do not contain, hydrogen. {The last quantity depends on
  the temperature and electron density and is computed in LTE}.

 Eq.~\ref{eq:edist} is solved
together with the equations for charge conservation, hydrogen nucleus
conservation, evolution equations for the 
atomic hydrogen level populations $n_i$ (5 bound levels plus the continuum):
\begin{equation} \label{eq:hevol}
 \frac{\partial n_i}{\partial t} + \nabla \cdot (n_i\vec{u}) =
 \sum_{j,j \ne i}^{n_\rml} n_j P_{ji} - n_i \sum_{j,j \ne i}^{n_\rml} P_{ij},
\end{equation}
{and an equation for non-equilibrium formation of \h2 molecules:}
\begin{equation} \label{eq:h2evol}
 \frac{\partial \nh2}{\partial t}  + \nabla \cdot (\nh2\vec{u})= R_\mathrm{f} n_1^3 -R_\mathrm{b}
 n_1 \nh2,
\end{equation}
where $n_1$ is the ground state population of atomic hydrogen.

The radiative part of the rate coefficients $P_{ij}$ is computed using
the approximations given by 
\citet{sollum1999},
{the temperature-dependent rate coefficients $R_\mathrm{f}$ and $R_\mathrm{b}$ are taken from the UMIST
database}
\citep[\url{www.udfa.net}]{2007A&A...466.1197W}. 

These equations yield values for $T$, $\nel$, $\nh2$, and
$n_i$. The gas pressure is then given by
\begin{equation} \label{eq:pressure}
P = k T \left( \nel + \nh2 + \sum_{i=1}^{n_\rml} n_i + \no \right),
\end{equation}
and used in the momentum and energy equations
(Eqs.~\ref{eq:momentum}--\ref{eq:energy}). {This non-equilibrium equation-of-state
requires advection of 6 atomic and 1 molecular hydrogen population in
addition to the 8 MHD variables, for a total of 15 advected
quantities.} See
\citet{2007A&A...473..625L}, 
\citet{golding2010}
and
\citet{Gudiksen++2011}
for more details.

\paragraph{Radiative heating.}
Formally, the radiative flux divergence is given by
\begin{equation} \label{eq:radfluxdiv}
\nabla \cdot \vec{F}_\mathrm{r} = \int_0^\infty \int_\Omega
 \alpha(\nu,\vec{\hat{n}}) \, \left( S(\nu,\vec{\hat{n}}) -
 I(\nu,\vec{\hat{n}}) \right) \, \rmd \Omega \, \rmd \nu,
\end{equation}
with $S$, $\alpha$ and $I$ the source function, opacity and intensity at
frequency $\nu$ in direction $\vec{\hat{n}}$ into the solid angle
$\Omega$. In practice, this double integral is too computationally
expensive to compute, and various approximations are made.

The integral over frequency is replaced by four
representative radiation bins, with their own associated bin-averaged
opacity per mass unit $\kappa_i$, photon destruction probability
$\epsilon_i$ and thermal emission $E_i$, assuming isotropic scattering and
isotropic source function and opacity
\citep{1982A&A...107....1N,2000ApJ...536..465S}. 
The radiative heating of such a so-called {\it multi-group} scheme is
then given by
\begin{equation}  \label{eq:qmultigroup}
Q_\mathrm{rad} =  4 \pi \rho \sum_{i=1}^4 \kappa_i (\epsilon_i J_i - E_i)
\end{equation}
The implementation of this
scheme in the Bifrost code is discussed in detail by
\citet{2010A&A...517A..49H}. 
The quantities $\kappa_i$, $\epsilon_i$ and $E_i$ are
precomputed and tabulated assuming LTE populations and line-scattering
in the 2-level Van Regemorter approximation. They are tabulated as
function of the electron density and gas temperature. This is an
improvement over previous work by
\citet{2007A&A...473..625L}
who used tables as function of the internal energy and the mass
density, implicitly assuming LTE values for the electron density and
the temperature. 


The multi-group scheme has the drawback that to properly approximate
the effects of the strongest spectral lines one needs to add several
bins just to cover the largest opacities. Otherwise the cooling and
heating effects of the few strong spectral lines in a bin otherwise
comprised of spectral features with much lower opacity are washed out
in the construction of the bin-averaged radiation
quantities (see \eg  \ Eq. 9 -- 11 of
\citet{2010A&A...517A..49H}).
Furthermore, the assumptions of an
opacity in LTE and a source function from the 2-level Van Regemorter
approximation do not work for the strongest chromospheric lines. In
the case of simulations of the chromosphere this means that the
radiative heating and cooling from the mid-chromosphere upwards is
badly modeled with a multi-group scheme only.

Therefore, the MHD model includes additional cooling in such lines in
the following way:

Optically thin radiative losses from the corona and transition region
are included through a frequency-integrated loss function $\Lambda(T)$
based on the coronal approximation for the level populations:
\begin{equation} \label{eq:qthin}
Q_\mathrm{thin} = - \Lambda(T)  \, \nel \,  n_\rmH.
\end{equation}
The function $\Lambda(T)$ is only significantly different from zero
above $T=15000$~K.

Optically thick parameterized radiative losses {and gains} from
\HI, \MgII\ and \CaII\ are included through:
\begin{equation} \label{eq:Qhca}
Q_\mathrm{[H,Ca,Mg]} =  C(T)_\mathrm{[H,Ca, Mg]} \, \nel \, \rho \,
 \min \left(\frac{\exp^{{-k_{\mathrm{[H,Ca,Mg]}}
  m_{\rm c}}}}{m_{\rm c}^{0.3}},1 \right).
\end{equation}
Here the constant k and the temperature-dependent coefficient C are
determined from detailed radiative transfer computations with the
RADYN
\citep[see \eg][]{1995ApJ...440L..29C} 
and Multi3D codes
\citep{2009ASPC..415...87L}, 
and $m_\rmc$ is the column mass. These functions include hydrogen lines and
the Lyman continuum and the pertinent lines and continua from \CaII\ and \MgII\
\citep[][in preparation]{Carlsson+Leenaarts2011}.
%

Radiative heating through absorption of coronal radiation in UV
continua is modeled through the representative bound-free absorption
cross-section $\sigma$ of \ion{He}{I} at 25~nm:
\begin{equation} \label{eq:QHe}
Q_\mathrm{He} = \sigma_\mathrm{He,25\,nm} \, n_\mathrm{He~I} \,
\exp^{-\tau_{\mathrm{He,25\,nm}}} \, J_\mathrm{thin}
\end{equation}
with $J_\mathrm{thin}$ the angle-averaged radiation field resulting from
$Q_\mathrm{thin}$
\citep[][in preparation]{Carlsson+Leenaarts2011}.
%

In addition we include an ad-hoc heating term that drives the
temperature up to $T_0=1660$~K once it drops below that value. It is given
by:
\begin{equation} \label{eq:Qwrm}
Q_\mathrm{w} = C_\mathrm{w} n_H^2 \left[\max \left( 0,(T_0-T) \right)\right]^2,
\end{equation}
with $C_\mathrm{w}$ a constant that sets the heating rate. The
simulation thus still has an artificially set minimum temperature, but
it is 740~K lower than in the simulation of
\citet{2007A&A...473..625L}.
{We add this ad-hoc term to prevent our simulation from cooling to
too low temperatures where the radiation tables we employ become
inaccurate.}

All the different contributions are added so that the radiative heating
term in Eq.~\ref{eq:energy} becomes
\begin{equation} \label{eq:qradtot}
- \nabla \cdot \vec{F}_\mathrm{r} = Q_\mathrm{rad} + Q_\mathrm{thin} +
Q_\mathrm{H} + Q_\mathrm{Ca} + Q_\mathrm{Mg} + Q_\mathrm{He} + Q_\mathrm{w}
\end{equation}

\paragraph{Simulation setup.}
We performed a 2D simulation with a grid size of $512 \times 325$,
spanning from 1.5~Mm below $\left< \tau_{500} \right> = 1$ to 14~Mm
above it in the lower corona. The horizontal grid spacing is 32.5~km,
the vertical grid spacing is 28~km from the convection zone up to the
low corona, and increased to 150~km higher up in the corona.

The lower boundary is open, allowing fluid to freely enter and leave
the box, while specifying the entropy of the inflowing gas to maintain
sufficient energy flux into the computational domain. The upper
boundary uses the methods of characteristics to extrapolate the MHD
variables, letting waves exit the domain with almost no
reflections. The upper boundary is set to strive to a temperature of
{250,000}~K to prevent the corona from cooling as a 2D weakly-magnetic
simulation cannot sustain a corona self-consistently. We choose
  this rather low temperature because we model a weakly magnetic part
  of the atmosphere. Its exact value has no influence on the behaviour
  of the chromosphere.  As formulated in the code, the thermal
conduction operator requires a magnetic field to be present, so a weak
magnetic field (average unsigned flux density in the photosphere of
0.3~G) is added, but is too weak to have any effect on the atmosphere.

The simulation was started from a previously relaxed snapshot computed
without non-equilibrium hydrogen ionization and ran for 1 hour of
solar time. We discard the first 10 minutes to remove start-up
effects.

\section{Results} \label{sec:results}

\begin{figure}
  \includegraphics[width=\columnwidth]{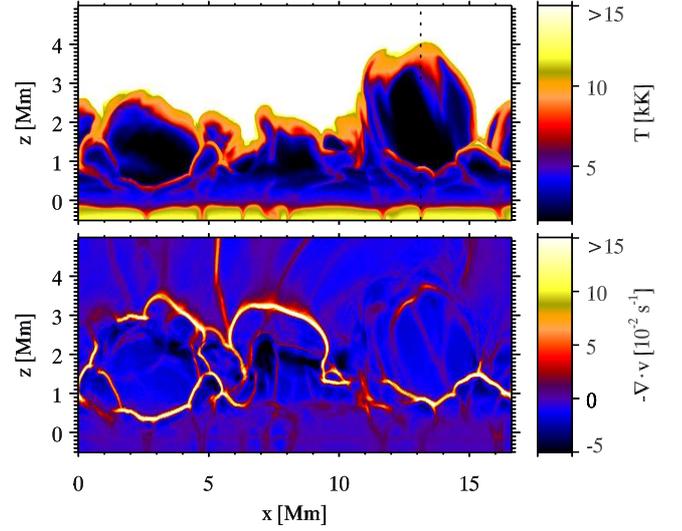}
  \caption{Snapshot of the simulated atmosphere. Top panel:
    gas temperature, with the column shown in Figs.~\ref{fig:atmos},
    \ref{fig:ebal} and \ref{fig:qgenrad} --- \ref{fig:chemeq}
    indicated by a dashed line. Bottom panel: $- \nabla \cdot
    \vec{u}$, positive values indicate the gas is locally compressed
    in the co-moving frame.
  \label{fig:atmosslice}}
\end{figure}

Figure~\ref{fig:atmosslice} shows the chromosphere in a snapshot from
the simulation after {34}~min of solar time have elapsed. The upper
panel shows the gas temperature, with granules visible below $z=0$~Mm,
the high-temperature transition region and corona in white at the
top. In between lies the chromosphere, visible as a cool background
state pervaded by waves and shock-fronts with peak temperatures
increasing with height. The bottom panel shows $- \nabla \cdot
\vec{u}$, a measure for the local compression rate of the
gas. Intergranular lanes and their extension into the upper
photosphere appear as weakly compressing purple stalks at the bottom
of the panel. The scene is dominated by strongly compressive shocks in
the chromosphere {and corona} with expanding shock-wakes, such as the one around
$(x,z) = (13,2)$~Mm. Note that some of the shock fronts are
co-spatial with the chromosphere-corona interface, and push the corona
upward. An extreme example of this is seen along the dashed line at
$x=13.1$~Mm in the upper panel. The shock front at $z=3.7$~Mm has
pushed the corona up, leaving a large pocket of cool
expanding gas in its wake. At $z=0.9$~Mm a new shock front is
propagating up through the cold gas, compressing it again.

\begin{figure*}
  \includegraphics[width=\textwidth]{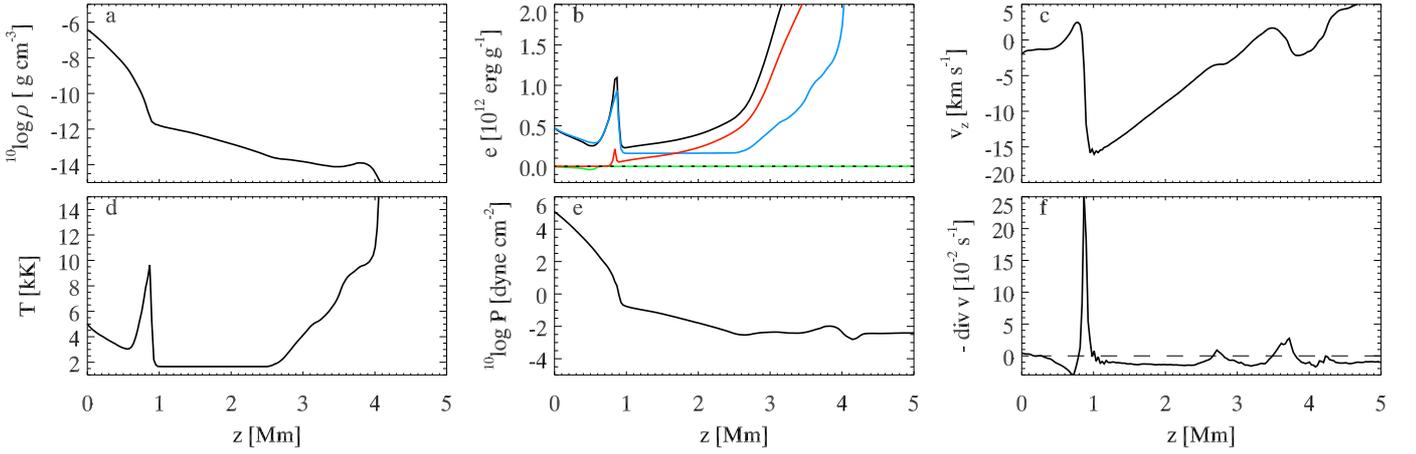}
  \caption{Properties of the atmosphere along the cut indicated in
    Fig.~\ref{fig:atmosslice}. a: Mass density; b: The solid black curve
    indicates the internal energy per mass unit, with the zero point
    at neutral atoms in the ground state. Red:
    ionization energy of hydrogen; black dotted: excitation energy of
    atomic hydrogen; blue: kinetic
    contribution of all atoms, molecules and electrons; green:
    contribution of the rotational, vibrational
    and dissociation energy of $\h2$ molecules. c: Vertical gas velocity,
    positive means upward. d: Gas temperature. e: Gas pressure. f:
    compression rate $- \nabla \cdot \vec{u}$.
  \label{fig:atmos}}
\end{figure*}

Figure~\ref{fig:atmos} shows various quantities along the cut
indicated by the dashed line in Fig.~\ref{fig:atmosslice}. \mbox{Panel a}
shows the density profile. It decreases with height everywhere except
around $z=3.8$~Mm at the site of the shock front that pushes the corona
upward. Panel~b shows the total internal energy in black with its
distribution over various contributions. It has a peak at the shock
front at $z=0.9$~Mm and a smooth increase with height above it. The
main contributors to the internal energy are the energy of the random
motions of the gas particles (blue) and the ionization energy of hydrogen (red).
The latter is the largest contribution above 1.6~Mm. There is a
small but important contribution of \h2 molecules at 0.5~Mm
height. Panel~c shows the vertical velocity. It exhibits a rough
sawtooth shape common to shock-propagation in a stratified
atmosphere. Panel~d shows the temperature, with a high-temperature
shock front at 0.9~Mm, and a plateau at 1660~K between 1 and
2.5~Mm. Panel e shows the gas pressure and panel f shows the
compression rate, again indicating the presence of the two shock
fronts at 0.9 and 3.8~Mm, weak compression at 2.8~Mm {and expansion in
the rest of the chromosphere.}

\begin{figure}
  \includegraphics[width=\columnwidth]{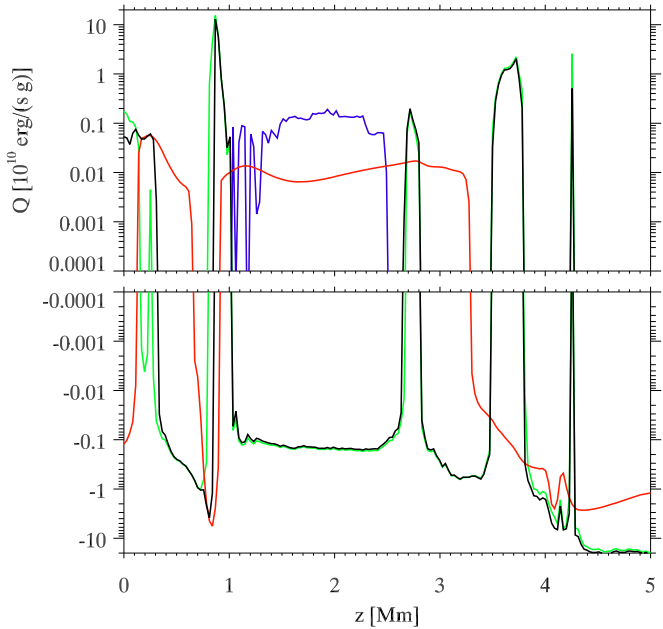}
  \caption{Energy balance along the cut indicated in
    Fig.~\ref{fig:atmosslice}. The upper part shows heating, the lower
    part cooling. Black: total heating rate (right-hand side of
    Eq.~\ref{eq:energy}, excluding the ad hoc term $Q_\rmW$. 
    Green: compression work
    $- P (\nabla \cdot \vec{u})$. Red: total radiative heating (Eq.~\ref{eq:qradtot}), excluding
    $Q_\rmW$. Blue: ad-hoc heating
    $Q_\rmW$  (Eq.~\ref{eq:Qwrm}).
  \label{fig:ebal}}
\end{figure}

Figure~\ref{fig:ebal} displays the energy balance in the chromosphere
(the right-hand-side of Eq.~\ref{eq:energy}) for the same column as
Fig.~\ref{fig:atmos}. Viscous energy dissipation, Joule heating and
thermal conduction are all negligible in the chromosphere and are not
shown. The energy balance is set by compression work and radiation. As
expected, the shock fronts are heated compressively and cool
radiatively. The cool area between 1 and 3.4~Mm is cooling through
expansion and is heated by radiation except at 2.7~Mm, where a
horizontally propagating wave causes some compression. The expansion
cooling is stronger than the radiation heating by an order of
magnitude. This imbalance ultimately leads to the ad-hoc $Q_\rmW$
contribution becoming active to prevent the atmosphere from cooling
below 1660~K. The instant in time shown in the figure shows a scene
with large ad-hoc heating. However, this term is only intermittently
active and switches off as soon as the temperature is above 1660~K.

\begin{figure}
  \includegraphics[width=\columnwidth]{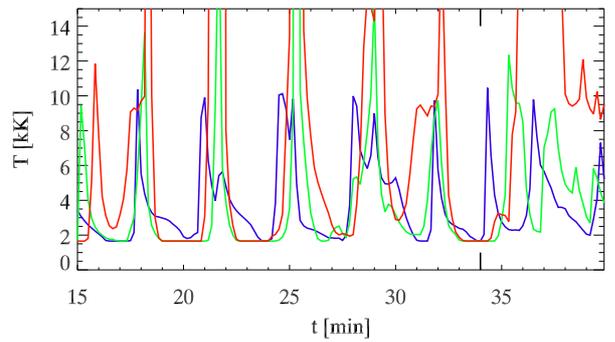}
  \caption{Time evolution of the temperature at different heights in
    the column indicated by the dashed line in
    Fig.~\ref{fig:atmosslice}.  The time moment shown in that figure
    is indicated by the long tick marks at $t=8.4$~min. Blue:
    $z=1$~Mm; green: 1.5~Mm; red: 2~Mm.
 \label{fig:time-evol}}
\end{figure}

Figure~\ref{fig:time-evol} shows the time evolution of the temperature
at different heights along the dashed line of
Fig.~\ref{fig:atmosslice}. The blue curve between $t=16$ and $t=24$ minutes
show a clean example of the temperature variations of passing shocks,
with a rapid increase in temperature and a more gradual cooling
phase after passage of the shock front. In general the time evolution
is more irregular because of slanted shock propagation and
interference. The corona intermittently dips down to below 1.5~Mm,
indicated by the red and green curves running off the temperature
scale.

\begin{figure}
  \includegraphics[width=\columnwidth]{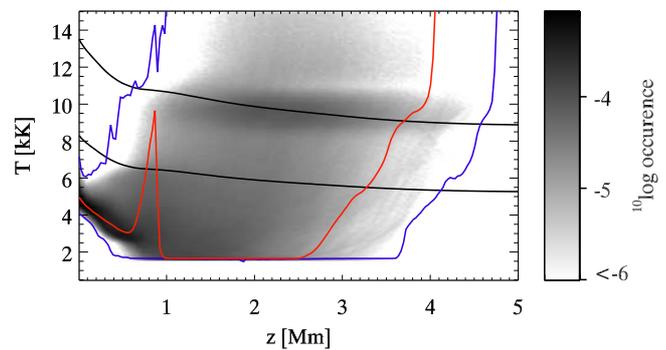}
  \caption{Diagram of the temperature occurrence as function of
    atmospheric height in the simulation. The red curve shows the
    temperature in the column marked in Fig.~\ref{fig:atmosslice}.
    The blue curves indicate the minimum and maximum temperatures as a
    function of height during the whole simulation run. The black
    curves shows the location of 50\% \ion{He}{i} (top curve) and
    \ion{H} (bottom curve) ionization assuming Saha equilibrium and
    the average run of the electron density with height.
  \label{fig:tghist}}
\end{figure}

The above discussion and Figs.~\ref{fig:atmos} -- \ref{fig:time-evol}
again confirm the picture of the quiet chromosphere as a layer that
undergoes quick compression during shock passages followed by a longer
phase of expansion cooling. 
The time scale of this expansion cooling $t_{P \rmd V} = e/ (P (\nabla
\cdot \vec{u}))$ varies from 200~s to 500~s assuming typical values of
$e=10^{12}$~erg~g$^{-1}$ and $2 \times 10^9 <  P \, (\nabla \cdot
\vec{u}) < 5 \times 10^9$ ~erg~s$^{-1}$~g$^{-1}$. Here $e$ is roughly
the thermal energy of solar gas at $12\,000$~K. The ionization energy
remains nearly constant, so the gas in the chromosphere behaves approximately
as an ideal gas. So, expansion cooling alone can, in cases of strong
expansion, cool chromospheric gas to very low temperatures between 2
shock passages, as the radiative heating rarely exceeds $2 \times
10^8$ ~erg~s$^{-1}$~g$^{-1}$ in the chromosphere.

This rough estimate is confirmed by Fig.~\ref{fig:tghist}.  It shows a
histogram of the occurrence of gas temperature values as a function
of atmospheric height. The range of temperatures increases with
increasing height when going up from the photosphere. The maximum
temperature curve shows the increase of peak shock temperature with
height up to 1~Mm. 
The maximum temperature at 1~Mm is above
$15\,000$~K, indicating that the corona occasionally reaches this far
down.
The thick dark band at $10\,000$~K between 1 and 4~Mm is
caused by a combination of shock fronts and the layer of $10\,000$~K
gas just below the corona (see the upper panel of
Fig.~\ref{fig:atmosslice}). This band is discussed in
Sec.~\ref{sec:hhetherm}. Between $z=0.8$~and~2.2~Mm the histogram
peaks along a narrow dark band below $T \approx 2000$~K, indicating
that such low temperatures are common in the simulated
chromosphere. Above 2.2~Mm such low temperatures occur less
frequently, but the atmosphere can be as cold as {the ad-hoc
  heating threshold temperature} of 1660~K up to 3.4~Mm height.

In the following subsections we investigate the accuracy of this result
by discussing the various physical processes that determine the
minimum temperature in the chromosphere and the accuracy with which
they are represented in the simulation.

\subsection{H$_2$ thermostat action}

\begin{figure}
  \includegraphics[width=\columnwidth]{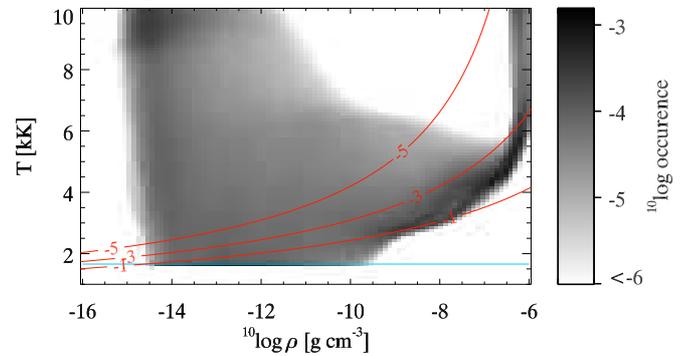}
  \caption{Diagram of the occurrence of temperature as function of mass density in the
    simulation. The red contours show the fraction of
    hydrogen atoms bound in \element{H}$_2$ molecules, $2 \nh2 /
    (n_\Hi+2 \nh2)$, as function of temperature and mass density,
    assuming ICE and all hydrogen neutral. The contours are spaced a factor 100
    apart, with the labels indicating the exponent $a$ in $10^a$. The
    blue line at $T = 1.66$~kK specifies the threshold for the
    ad hoc heating (Eq.~\ref{eq:Qwrm}).
  \label{fig:fh2}}
\end{figure}

When the low-chromospheric temperature reaches down to 2000--3000~K, \h2
molecules can form in significant amounts. This process is exothermic,
releasing 4.48~eV per molecule formed, compared to a thermal energy of
$kT=0.19$~eV per particle at 2000~K. Thus, as \h2 begins to form it
releases a large amount of energy, which by Eq.~\ref{eq:edist} is
predominantly converted to thermal energy, counteracting the expansion
cooling. This effect is illustrated in Fig.~\ref{fig:fh2}. It shows
the simultaneous occurrence rate of combinations of temperature and
mass density in the simulation, with overplotted contours of the
fraction of hydrogen atoms bound in \h2 {assuming instantaneous
chemical equilibrium.}

The minimum temperature in the chromosphere follows the curve that
represents 10\% of all hydrogen bound in \h2 between $\rho =
10^{-9}$~g~cm$^{-3}$ and $10^{-7.5}$~g~cm$^{-3}$. At lower mass
densities (higher up in the chromosphere) {the formation rate of
  \h2 becomes so low that it cannot form in large enough amount to
  prevent the expanding parts of the atmosphere from cooling
  further. This is indicated by the turnoff of the bottom of the grey
  cloud away from the red curve at $\rho = 10^{-9}$~g~cm$^{-3}$. Once
  the temperature drops below 1660~K the ad-hoc} heating
becomes active, preventing the atmosphere from cooling even further.

\subsection{\ion{H}{} and \ion{He}{} as
  thermostats}  \label{sec:hhetherm}

Figure~\ref{fig:atmosslice} shows a layer of $10\,000$~K hanging as a
ragged skirt below the corona. This layer is one of the causes of the
dark band at the same temperature in Fig.~\ref{fig:tghist}.  The
overlaid \ion{He}{} ionization curve demonstrates that \ion{He}{i}
works similarly to \h2 as a thermostat in this simulation.  This is
unphysical, however, since the actual ionization-recombination balancing
is likely to be even more out of instantaneous equilibrium for helium
than in the case of hydrogen. Thus, the Saha equilibrium
assumed in our simulation for helium is erroneous.  Relaxing this
assumption is likely to remove helium's thermostat action, just as for
non-equilibrium hydrogen ionization
\citep{2002ApJ...572..626C,2007A&A...473..625L}.
Indeed, hydrogen does not cause any such thermostat clustering of
temperature values (which would be at the lower black curve in
Fig.~\ref{fig:tghist}) since it is treated properly in
non-equilibrium.  The erroneous assumption of LTE balancing for
\ion{He}{i} does not affect the present analysis, however, since it
affects only the hottest phases of the chromospheric gas and not the
coolest ones addressed here.

\subsection{Radiative heating} \label{sec:radheat}

\begin{figure}
  \includegraphics[width=\columnwidth]{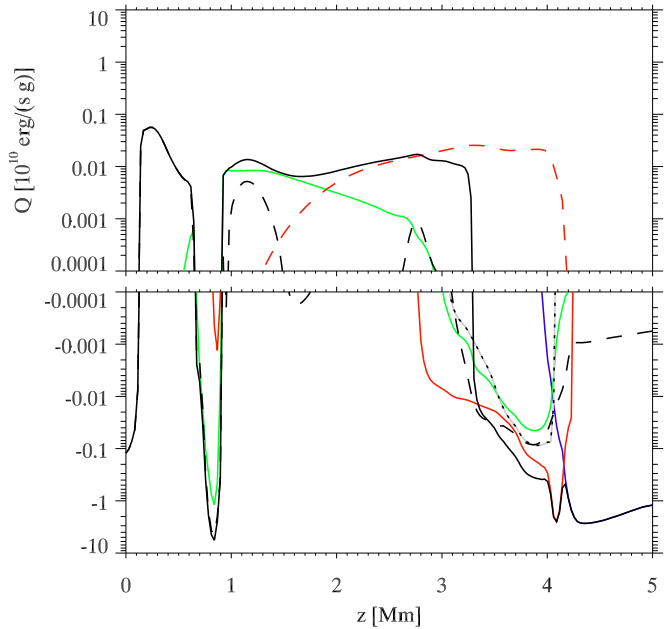}
  \caption{Various contributions to the total radiative heating rate
    (Eq.~\ref{eq:qradtot}) for the column indicated in
    Fig.~\ref{fig:atmosslice}. Black solid: total radiative
    heating. Black dashed: Radiative heating from the multi-group
    scheme ($Q_\mathrm{rad}$, see Eq.~\ref{eq:qmultigroup}). Red:
    $Q_\mathrm{H}$. Green: $Q_\mathrm{Ca}$. Blue:
    $Q_\mathrm{thin}$. Red dashed: $Q_\mathrm{He}$. Grey with black dots:
    $Q_\mathrm{Mg}$.
  \label{fig:qgenrad}}
\end{figure}

The chromosphere is heated by radiation during its cool
phases. Fig.~\ref{fig:qgenrad} shows the various contributions that
make up the total radiative heating rate in our simulation along the
column indicated in Fig.~\ref{fig:atmosslice}.

The radiative heating due to the multi-group scheme
(Eq.~\ref{eq:qmultigroup}) dominates the heating below $z = 1$~Mm as
indicated by the near-equality of the solid and dashed black curves. It
also cools strongly in the upper chromosphere. Absorption of coronal
radiation (red dashed curve, Eq.~\ref{eq:QHe}) adds heating above 1.3~Mm
and is dominant between 1.8 and 3.2~Mm. The \CaII\ lines (green,
Eq.~\ref{eq:Qhca}) cool in the shock front at 0.9~Mm and above 2.8~Mm,
they are the dominant contribution in the cool gas between 1 and
1.8~Mm. Hydrogen (red, Eq.~\ref{eq:Qhca}) cools in the shock front at
0.9~Mm and above 2.8~Mm, where the Lyman lines and continua become
effectively optically thin. Optically thin losses from the corona
(blue) do not play a role in the chromosphere. The \ion{Mg}{ii}
(grey-black dotted) lines only cool significantly in the upper
chromosphere.

We will now discuss the different contributions in detail and show
them for the representative column from our simulation displayed in
Figs.~\ref{fig:atmos} -- \ref{fig:time-evol}.

\paragraph{Negative hydrogen ion \Hmin.}

\begin{figure}
  \includegraphics[width=\columnwidth]{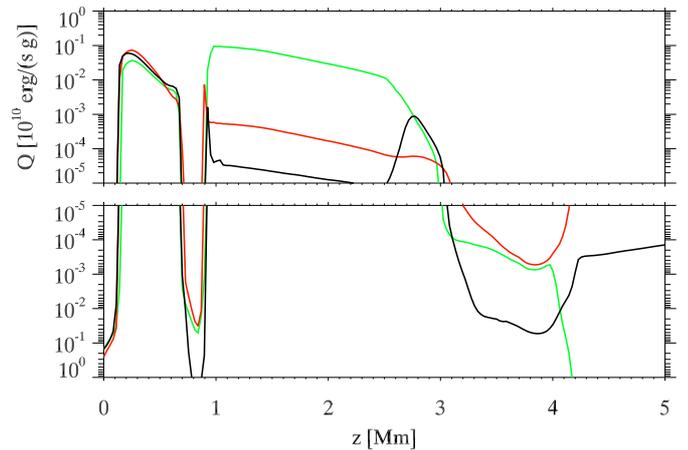}
  \caption{Comparison of the multi-group radiative heating rate in the first bin
    (black solid, $Q_1$ see Eq.~\ref{eq:qmultigroup}) compared to
     the heating rate due to \Hmin\ in NLTE (red) and LTE (green).
  \label{fig:qhminrates}}
\end{figure}

Absorption of radiation in the \Hmin\ continuum is a potentially large
source of heating. This absorption is included in the first radiation
bin of our model. We computed the \Hmin\ heating in detail for the
snapshot of Fig.~\ref{fig:atmosslice} assuming LTE, \ie\ the
source function is the Planck function and the \Hmin\ density is set by
its Saha-equilibrium with neutral hydrogen. For comparison, we computed
the \Hmin heating rate in NLTE including the following reactions:
\begin{eqnarray}
\Hmin + h \nu & \leftrightarrow & \rmH + e, \nonumber \\
\Hmin + H &\leftrightarrow & 2 \rmH + e,    \nonumber \\
\Hmin + e & \leftrightarrow & \rmH + 2e,    \nonumber \\
\Hmin + p & \leftrightarrow & 2 \rmH,       \nonumber \\
\Hmin + H & \leftrightarrow & \rmH_2 + e,   \nonumber 
\end{eqnarray}
where we kept the neutral hydrogen, proton, electron and,
\h2 density constant. This constancy is justified because of the high
number density of these particles relative to the \Hmin density.
The reaction rates for the latter four reactions are from
\citet{1968MNRAS.141..299L}.

Figure~\ref{fig:qhminrates} shows a comparison with the first bin of
the multi-group heating rate (black, Eq.~\ref{eq:qmultigroup} and the
heating rate caused by \Hmin bound-free and free-free transitions,
assuming LTE (green) and NLTE (red). The LTE assumption generates the
largest heating, caused by both the low Planck function compared to
the average radiation field, and the high \Hmin number density. The
NLTE heating rate is much smaller. This is caused by a combination of
photo-ionization of \Hmin, leading to a NLTE departure coefficient
much smaller than unity, and strong scattering, leading to smaller
$J_\nu - S_\nu$ splitting. Absorption of radiation by \Hmin\ is
therefore not efficient in heating the chromosphere.

The first bin in the multi-group scheme closely matches the heating
due to \Hmin\ where \Hmin\ dominates the opacity (up to
0.8~Mm). {It deviates in the shock around 0.9~Mm and above. The
  first bin underestimates the NLTE \Hmin\ heating in the cool area
  between 1 and 2.5~Mm. Above 3~Mm it erroneously overestimates the
  NLTE cooling by several orders of magnitude.}

\paragraph{\CaII.}

\begin{figure}
  \includegraphics[width=\columnwidth]{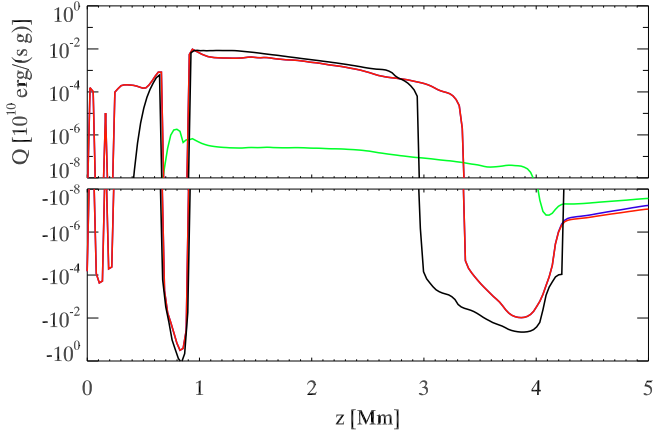}
  \caption{Comparison of the parameterized radiative heating rate due to
    \CaII\ ($Q_\mathrm{Ca}$, see Eq.~\ref{eq:Qhca}) to the
    heating computed based on a detailed radiative transfer computation.
     Black: $Q_\mathrm{Ca}$. Blue: heating due to
    \CaII\ lines. Green: heating due to \CaII\ continua. Red: total
    \CaII\ heating. The blue curve is nearly equal to the red curve.
  \label{fig:qcarates}}
\end{figure}

The lines of \CaII\ are other strong heating agents in the
chromosphere. We computed the \CaII\ heating rates for the snapshot of
Fig.~\ref{fig:atmosslice} using the radiative transfer code
Multi3d. This detailed computation was performed {treating each
  column in the simulation as a plane-parallel atmosphere} and
included the effects of partial redistribution (PRD) in the
\HK\ lines. The results for the representative column are shown in
Fig.~\ref{fig:qcarates}. The \CaII\ continua (green curve) have a
negligible effect on the total heating rate. The lines cool strongly
in the shock front at 0.8~Mm and above 3.4~Mm. They heat the
atmosphere in the cool area between 0.9 and 3.4~Mm.

The parameterized \CaII\ cooling employed in our simulation is in
reasonable agreement with the detailed computation above 0.5~Mm
height. The largest differences are the overestimation of the cooling
between 3.5 and 4.2~Mm and shifted
location around 3~Mm where the heating term switches sign.
The difference below 0.5~Mm is caused by a cutoff to avoid doubling the
heating rate as heating at low heights is also included in the
multi-group scheme.

Manual inspection of the heating rate in many different columns of the
simulation shows that the parameterized heating is accurate most of
the time.

\paragraph{\MgII.}

\begin{figure}
  \includegraphics[width=\columnwidth]{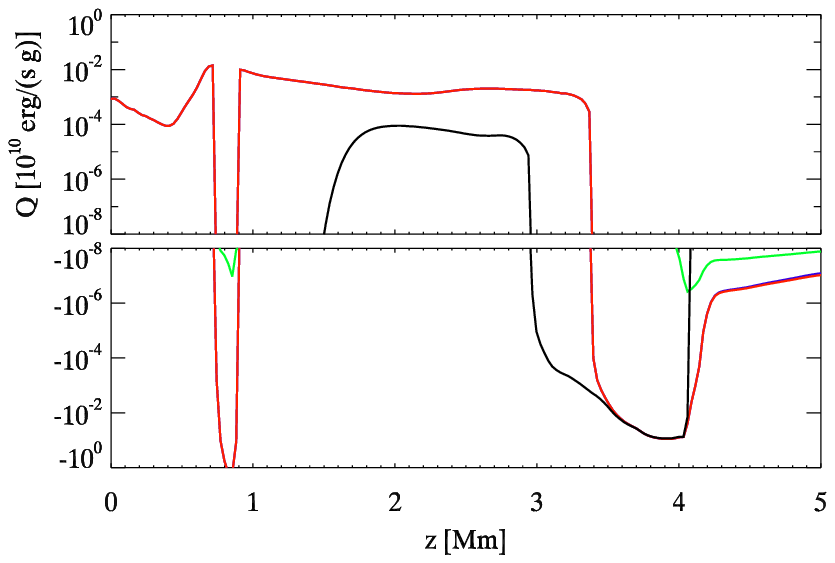}
  \caption{Comparison of the parameterized radiative heating rate due to
    \ion{Mg}{ii} ($Q_\mathrm{Mg}$, see Eq.~\ref{eq:Qhca}) to the
    losses computed based on a detailed radiative transfer computation.
   Black: $Q_\mathrm{Mg}$. Blue: heating due to
    \MgII\ lines. Green: heating due to \MgII\ continua. Red: total
    \MgII\ heating. The blue curve is nearly equal to the red curve.
  \label{fig:qmgrates}}
\end{figure}

Another potentially large contributor to the total chromospheric
heating is \MgII. We computed the radiative losses in detail including
the effects of PRD in the h\&k lines. The result for the
representative column is given in Fig.~\ref{fig:qmgrates}. The
 h\&k lines cool in the shock front at 0.9~Mm, heat the cool
area above the shock front and cool the upper chromosphere. The
continua play a minor role.

{The cutoff in the parameterized heating pushes the heating to
  zero at 1.5~Mm. The heating is much lower than the detailed
  computation in the cool area between 1.5 and 3~Mm and it shows the
  same shift as for \CaII\ in the location where the heating switches
  sign.} The cooling peak around 3.8~Mm is reproduced correctly by
$Q_\mathrm{Mg}$.

Manual inspection shows that the parameterized \MgII\ heating is
accurate for gas temperatures above 6000~K. Below this temperature the
heating rate is modeled qualitatively correct, but can deviate because
the effects of the precise atmospheric structure, in particular the
velocity field are not taken into account in the simple parameterization.

\paragraph{Hydrogen.}

\begin{figure}
  \includegraphics[width=\columnwidth]{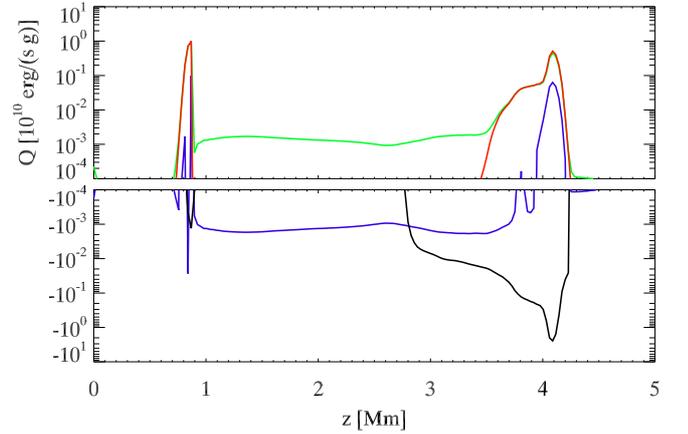}
  \caption{Comparison of the parameterized radiative heating rate due
    to hydrogen ($Q_\mathrm{H}$, see Eq.~\ref{eq:Qhca}) to the losses
    implied by the non-equilibrium hydrogen radiative rates. Black:
    $Q_\mathrm{H}$. Blue: implied heating due to hydrogen
    lines. Green: implied heating due to hydrogen continua. Red: total
    implied heating.
  \label{fig:qhrates}}
\end{figure}

Hydrogen has only a small effect on the heating rate of the lower and
middle chromosphere in the semi-empirical time-independent VAL C
model atmosphere
\citep{1981ApJS...45..635V}. 
In this model the \Lyalpha\ transition is so optically thick that it
does not play a role in the energy balance, and the losses in
\Halpha\ are approximately offset by the absorption in the Balmer
continuum. This effect remains valid in the time-dependent models of
\citet{2002ApJ...572..626C} 
computed with the RADYN code, except in shock fronts.  Unfortunately
it is currently not possible to compute the detailed time-dependent
heating rate due to hydrogen in multi-dimensional simulations, so we
cannot rigorously prove the same effect holds in our 2D simulation.
However, the small hydrogen heating rate is mostly an effect of the
atomic structure of hydrogen, and does not depend strongly on the exact
chromospheric structure. Therefore, our 2D simulation should behave
in a similar manner. This is confirmed in Fig.~\ref{fig:qhrates}. 

This figure shows the parameterized heating due to hydrogen (Eq.~\ref{eq:Qhca})
in black. The shock front at 0.9~Mm is cooling, there is little heating
between 1.2 and 2.5~Mm, and slowly increasing cooling above 2.8~Mm,
peaking at 4.1~Mm where the \Lyalpha\ line becomes optically thin.

Figure.~\ref{fig:qhrates} also shows the heating implied by the
radiative transition rates computed from Eq.~\ref{eq:hevol}.  We
define the implied heating rate in a transition between levels $i$ and
$j$ ($i<j$) by
\begin{equation}
Q_{\mathrm{imp}} = h \nu_0 \left(R_{ij} - R_{ji} \right), 
\end{equation}
with $\nu_0$ the line center or ionization edge frequency, and
$R_{ij}$ the radiative rate from level $i$ to level $j$. This simple
definition makes an error for bound-free transitions due to the
non-negligible width of the ionization edges, but this error is
small. Note that the implied heating does not appear as a source term
in Eq.~\ref{eq:qradtot}, the assumed radiation field only serves to
define the radiative rates in Eq.~\ref{eq:hevol}. It reproduces the
near-zero heating in the cool area between 1~Mm and 2.8~Mm. The
shock-front and the upper chromosphere are heated.

We conclude that both the parameterized cooling and the implied
heating rate reproduce the lack of hydrogen heating in the cool phases
of the chromosphere. Any errors caused be the approximations have
negligible effect on the minimum temperature in our model due to the
small heating rate compared to other heating agents.

\paragraph{Absorption of coronal radiation.}

Nearly all coronal radiation emitted towards the chromosphere is
eventually absorbed (a small part is backscattered into
space). Exactly where this radiation is absorbed depends on its
detailed spectral energy distribution and the absorption coefficient
in the chromosphere. However, these details are of minor importance to
the overall heating rate as long as all energy is absorbed. Our
simulation correctly models this behavior. The finite extent of the
corona in the simulation and its relatively low temperature might
result in a too low amount of radiation absorbed in the
chromosphere. {The coronal loss function ($Q_\mathrm{thin}$,
Eq.~\ref{eq:qthin}) peaks strongly in the lower corona due to its
quadratic density dependence, so we expect this error to be small.}

\section{Discussion and conclusions} \label{sec:discussion}

\begin{figure}
  \includegraphics[width=\columnwidth]{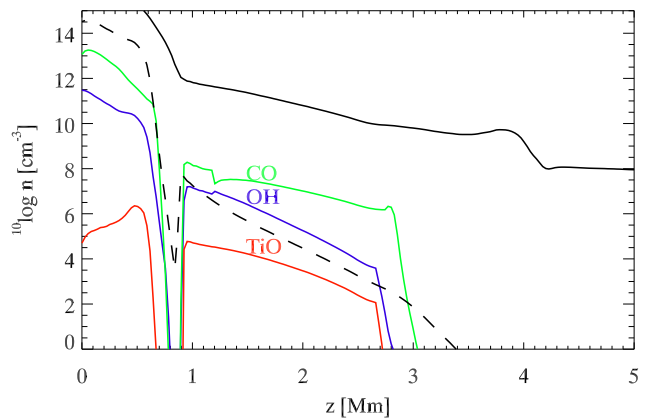}
  \caption{{Molecular densities in the exemplary column indicated
      by a dashed line in Fig.~\ref{fig:atmosslice}. The densities of
      CO (green), OH (blue) and TiO (red) are based on instantaneous
      chemical equilibrium. The \h2 density (black dashed) is based on
      our EOS. The total number density of hydrogen atoms is plotted
      in solid black for comparison. }
  \label{fig:chemeq}}
\end{figure}

\paragraph{Robustness of the minimum temperature}

{The formation of \h2 acts as a thermostat in the lower
chromosphere, preventing the temperature from dropping below
2000~K. ICE is not valid in the middle and upper chromosphere. There,
the slow formation rate of \h2 prevents thermostat action,
allowing solar gas to cool well below 2000~K. Our simulation correctly
models this behavior.}

In Sec.~\ref{sec:radheat} we discussed the various lines and
bound-free transitions
that can radiatively heat the chromosphere. We showed heating in lines
of \CaII\ and absorption of coronal radiation are the dominant
processes. We showed that \Hmin\ is inefficient in heating cool
pockets of chromospheric gas owing to strong scattering and low \Hmin
population due to photo-ionization. \MgII\ and hydrogen play a minor
role in the heating of the chromosphere. Our simulation accurately
models the heating due to \CaII\ and coronal radiation. The other
processes are less accurately modeled but these have, in sum, only a
small effect on the energy balance.

{The heating due to acoustic waves is self-consistently included in the 
simulation since we include the upper convection zone and the stochastic 
excitation of acoustic waves there. The limited resolution may lead to an 
underestimate of the excitation of high-frequency waves. A wave at a 
frequency of 20~mHz is represented with 11 grid-points so the effect will only be
important for really high frequency waves where simulations and observations 
indicate the effect for the heating of the chromosphere is minimal
\citep{2005Natur.435..919F,2007PASJ...59S.663C}.
}

The ad-hoc heating term acts intermittently in the largest
expansion bubbles, thus keeping our minimum chromospheric temperature
at 1660~K.

{We do not expect that a 3D simulation with similar physics will
  change this result. Three-dimensional simulations, such as the one
  reported on by
\citet{2010ApJ...709.1362L}
show temperatures down to 2000~K even with LTE hydrogen ionization
and instantaneous \h2 formation. It is very likely that such 3D
simulations with a non-equilibrium EOS would show the
same low temperatures as reported here (see Fig.~4 of
\citet{2007A&A...473..625L}
for an illustration of the resulting temperature differences between
these equations-of-state).}

We therefore conclude that our simulation provides an upper
bound to the minimum temperature of the non-magnetic chromosphere. The
simulation predicts the occurrence of gas temperatures as
low as 1660~K up to 3.4~Mm height. This low temperature is common
between 0.7 and 2.5~Mm height. Our ad-hoc heating prevents our model
from cooling further. This means that a non-magnetic chromosphere can
have temperatures even lower than this value. It seems impossible to
avoid such low temperatures using only radiation and hydrodynamic
processes.

\paragraph{Is the chromosphere really this cool?}

The sun has a turbulent photospheric magnetic field of the
order of 130~G
\citep{2004Natur.430..326T}. 
This turbulent field is likely to extend into the quiet chromosphere and
might generate enough Joule heating to prevent the quiet chromosphere
from cooling down to temperatures as low as in our simulation. In
addition the presence of a relatively strong magnetic field will also
change the propagation properties of the generated acoustic waves,
reducing their amplitudes by confining them to flux tubes and/or by
conversion to other magneto-hydrodynamic modes in the vicinity of the
$\beta=1$ surface. 

The
simulations of a solar surface dynamo by
\citet{2007A&A...465L..43V}
and
\citet{2008A&A...481L...5S} 
suggest that the strength of the turbulent field declines rapidly
with height, which would limit the amount of mid and high
chromospheric heating the field could cause. On the other hand,
the work of
\citet{2003ApJ...597L.165S}
suggests that part of the network magnetic fields can connect back
in the internetwork photosphere. Such a multiscale magnetic carpet
would lead to an increase of the field strength in the quiet chromosphere
relative to a purely locally generated field, with a corresponding
increase in the potential for Joule heating. 

Magnetic fields in the low ionization-degree
chromosphere may also lead to the generation of electric currents
through neutral-ion drag
\citep{2010ApJ...724.1542K}.  

Our simulation serves as a baseline case to test the effect of such
magnetic heating processes. It should be compared against 3D models like the one
by
\citet{2007A&A...465L..43V} 
but extended up into the corona, and including all the physics
discussed in this paper to study the effect of the turbulent
field. Such a simulation does not require a large horizontal extent of
the computational domain and, {while computationally very
  demanding, can in principle be done on the
  largest currently available supercomputers}.

Simulation of the effect of the magnetic carpet requires a larger
horizontal extent to include some network magnetic field and a large
enough area of quiet sun. The high spatial resolution ($\approx 10$~km
grid spacing) needed to get local dynamo action combined with the
large size of a network cell ($\approx 30$~Mm) makes such a simulation
challenging.

At this moment the minimum temperature of the quiet sun chromosphere
remains an open question. {Our simulation shows a non-magnetic
  chromosphere will get colder than our ad-hoc limit of 1660~K, but it
  is unknown whether regions uninfluenced by magnetic fields occur in
  the chromosphere. However, if these low temperature areas exist,
  they can in principle be observed.}

\paragraph{Suggestions for observations}
Direct observation of cool pockets of chromospheric gas is
hard. Atomic spectral lines with sufficient opacity form generally in
NLTE, making it hard to derive temperatures. The Atacama Large
Millimeter Array (ALMA) would be the ideal instrument to probe
chromospheric temperatures because of its high resolution, the LTE
source function and relatively well-defined formation height range of
the sub-millimeter continua
\citep{2004A&A...419..747L,2007A&A...471..977W}.

Alternatively, off-limb spectroscopy in molecular lines with
sub-arc-second resolution and low stray light could be used. As an
example we show the number densities of some abundant molecules as a
function of height in our {exemplary column in
  Fig.~\ref{fig:chemeq}. Because the chemical reaction timescales
  become very large in the chromosphere the number densities for the
  species computed assuming ICE} should be taken as a maximum possible
value only, and are likely orders of magnitude too high. The molecules
of choice for such observations seem to be \h2\ and CO. The solar
spectrum shows \h2\ UV emission lines
\citep{1978ApJ...226..687J}
that can be observed by the SUMER instrument aboard the Solar and
Heliospheric Observatory (SOHO) spacecraft
\citep{2006ASPC..354..230K}.
The mere presence of molecular lines at several arc-seconds above
the limb would prove the existence of low temperature gas at
high-chromospheric heights. Unfortunately the reverse is not true, the
absence of lines might merely indicate the absence of molecules and
not the absence of cool gas.

\paragraph{Conclusions} We have performed a 2D radiation-MHD simulation of
the non-magnetic solar chromosphere, including the effects of
non-equilibrium ionization of hydrogen and {non-equilibrium
  formation of \h2 molecules} and NLTE radiative transfer. We find
that our simulation contains pockets of cool gas with temperatures
{down to $1660$~K from 0.8~Mm up} to 3.4~Mm height. We discuss the physical
mechanisms that set the minimum temperature and find that our
simulation likely overestimates the minimum temperature. We conclude
it is impossible to avoid such low temperatures with
radiation-hydrodynamical processes only.

Such cool pockets of chromospheric gas might be even cooler in the
real sun than in our simulation, provided a quiet chromosphere without
significant magnetic heating exists. We suggest off-limb molecular
spectroscopy to look for such pockets, and suggest 3D simulations with
sufficient resolution to support a local dynamo with and without
network-like magnetic fields to investigate whether the assumption of
negligible magnetic heating is justified.

\begin{acknowledgements}
 We thank N.~Vitas for compiling data on
  molecular partition functions and chemical equilibrium constants. We
  thank R.~J.~Rutten for illuminating discussions and improvements to
  the manuscript.
   JL recognizes support from the Netherlands Organization for
  Scientific Research (NWO).
This research was supported by the Research Council of Norway through
 the grant ``Solar Atmospheric Modelling'' and 
 through grants of computing time from the Programme for Supercomputing.
\end{acknowledgements}

\bibliographystyle{aa} 
\bibliography{%
abbett,%
ayres,%
carlsson,%
golding,%
hyman,%
jordan-carole,%
kuhn,%
lambert,%
leenaarts,%
noyes,%
nordlund,%
rutten,%
schrijver,%
skartlien,%
sollum,%
temp,%
trujillo-bueno,%
vernazza,%
voegler,%
wedemeyer,%
wedemeyer-boehm%
}

\end{document}